\documentclass[twocolumn,useAMS,usenatbib,usegraphicx]{mn2e}

\newcommand{\cardiff}{{School of Physics \& Astronomy, Cardiff University, 5 The Parade, Cardiff, CF24 3AA, Wales, United Kingdom}}

\title[Clustering of FIR/sub-mm Selected Galaxies]{Models for the Clustering of Far-Infrared and Sub-millimetre selected Galaxies}

\author[Short \& Coles]{Jo Short and Peter Coles\\ \cardiff \\}

\begin{document}

\maketitle

\begin{abstract}
We discuss and compare two alternative models for the two-point angular correlation function of galaxies detected through the sub-millimetre emission using the \emph{Herschel} Space Observatory. The first, now-standard \emph{Halo Model}, which represents the
angular correlations as arising from one-halo and two-halo contributions, is flexible but complex and rather unwieldy. The second model is based on a much simpler approach: we incorporate a fitting function method to estimate the \emph{matter} correlation function with approximate model  of the bias inferred from the estimated redshift distribution to find the \emph{galaxy} angular correlation function. We find that both models give a good account of the shape of the correlation functions obtained from published preliminary studies of the HerMES and H-ATLAS surveys performed using \emph{Herschel}, and yield consistent estimates of the minimum halo mass within which the sub-millimetre galaxies must reside. We note also that both models predict an inflection in the correlation function at intermediate angular scales, so the presence of the feature in the measured correlation function does not unambiguously indicate the presence of intra-halo correlations. The primary barrier to more detailed interpretation of these clustering measurements lies in the substantial uncertainty surrounding the redshift distribution of the sources.
\end{abstract}

\section{Introduction}
\label{SECintro}
Current theoretical models predict that galaxies form and evolve in cold dark matter (CDM) halos. Galaxies consequently tend to trace the distribution of mass, although the manner in which they do this may be {\em biased} \citep{Kaiser1984,BBKS,Coles1993,Mo1996}. In principle, therefore, once the bias is allowed for, it is possible to use measurements of the clustering of galaxies to determine the clustering properties of the dark matter, especially if measurements can be made as a function of redshift. Because clustering evolution is sensitive to the parameters underlying the background cosmological model, such observations can provide another (independent) test of the concordance cosmological model; see, e.g. \cite{Coles2005}. In addition, clustering observations can be used to constrain properties of the galaxies themselves, such as the minimum halo mass within which they reside, which may yield clues about the processes of galaxy formation and evolution.

A steadily increasing number of surveys of large-scale galaxy clustering are now becoming available. In the optical wavebands there are projects such as the UKIDSS Ultra Deep Survey \citep{Hartley2010} and the SDSS Redshift Survey \citep{Connolly2010, Ross2009} which are being used to extract information on clustering as a function of redshift. The \emph{Herschel} Space Observatory was launched in 2009 and is the only space observatory to cover a spectral range from the far infrared to sub-millimetre and therefore provides a new and unique window through which to study high-redshift galaxy clustering. Two surveys of particular interest to this article, HerMES \citep{Oliver2010} and H-ATLAS \citep{Eales2010}, have already released angular correlation results \citep{Cooray2010,Maddox2010} from their Science Demonstration Phase (SDP) which we will discuss further later, so it is timely to raise the issue of modelling sub-millimetre galaxy clustering.

The so-called Halo Model has been used extensively over the past few years in modelling galaxy clustering in a variety of contexts. The
Halo Model combines approximations of the dark matter profile within individual halos, the mass function, and bias models to estimate the
correlation function for given cosmological parameters and characteristic halo masses. It has been shown to provide accurate and reliable predictions of clustering measurements, at the price of a certain degree of complexity and modelling freedom.

The main focus of this paper is a comparison of the Halo Model and a fitting function method which was initially introduced by \cite{Hamilton1991}, and subsequently developed by \cite{Peacock1994}, \cite{JMW95} and especially \cite{Peacock1996}, to estimate the matter correlation function. Work by \cite{Matarrese97} followed by \cite{Moscardini1998} and \cite{Coles1998}, showed how to incorporate this idea into a technique for providing detailed predictions of angular correlation in high-redshift galaxy surveys. In this paper we compare the predictions of this much simpler approach with results from the Halo Model.

The paper is organized as follows. In the next section, we outline the methodology for the fitting function method. Following that in
Section \ref{SECpoint2}, the models are compared to recent \emph{Herschel} data and the best-fitting values of the free parameters are extracted and compared. In Section \ref{SECcon}, we conclude with a summary.

\section{Models of galaxy clustering}
\label{SECmodels}
A model for the angular correlation function of galaxies basically involves a model for the three-dimensional clustering of galaxies as
a function of epoch (i.e. redshift $z$) combined with a machinery (derived from a model of the background cosmology) for projecting this information down the observer's light cone.

\subsection{Angular correlation function}
The observed angular correlation function, $\omega_\textsuperscript{obs} (\theta)$, is calculated from the spatial power spectrum, $\Delta^2_{\rm g}(k,z)$, using:
\begin{eqnarray}
  \omega_\textsuperscript{obs} (\theta) = \int\limits_Z G(z) \mathcal{N}^2(z)  \int\limits_{-\infty}^\infty \xi\left(r(u,\theta,z),z\right) du dz,
\end{eqnarray}
where $\mathcal{N}(z)$ is the normalized distribution of galaxies over redshift, $G(z) = (dx/dz)^{-1}$ where $x(z)$ is the comoving distance to redshift $z$, the separation $r(u,\theta,z) = \sqrt{u^2 + x^2(z)\theta^2}$ Mpc$h^{-1}$, and $\xi(r,z)$ is the two-point (galaxy-galaxy) correlation function where,
\begin{eqnarray}
\xi(r,z) = \int\limits^\infty_0 \Delta_{\rm g}^2(k,z) \frac{\sin(kr)}{kr}\frac{dk}{k}.
\end{eqnarray}
The galaxy power spectrum can be approximated using different methods such as those described below.

\subsection{Power Spectrum: Halo Model}
The Halo Model estimate for the galaxy power spectrum represents it as arising from two distinct components:
\begin{equation}
\Delta_{\rm g}^2(k,z) = \Delta_{1h}^2(k,z)+\Delta_{2h}^2(k,z),
\end{equation}
The first term represents contributions from galaxies residing within the same dark matter halo, while the second is generated by contributions from galaxies in separate halos. The one-halo term models small-scale highly non-linear clustering using a recipe that involves a number of different components, including the dark matter profile of the halos, the mass function of halos, and the halo occupation distribution (HOD); the two-halo term depends on the underlying matter power-spectrum as well as the bias of galaxy clustering relative to mass clustering. In this paper the dark matter halo density profile is approximated using the NFW profile suggested by \cite{Navarro1997}. The mass function is effectively the number density of halos at a given mass and for this we use the approximation found by \cite{Sheth1999}. The Halo Occupation Distribution (HOD) is a simple way of relating the distribution of galaxies from the dark matter distribution, by estimating how many galaxies are expected within a dark matter halo of given mass. This is split into two estimates - the number of central galaxies (which would either be 1 or 0) and the number of satellite galaxies; a summary of the procedure can be found in \cite{Ross2009}. There are extensions to the model (e.g. \citealt{Gil-Marin2010}) but we just consider the basic version here. \cite{Cooray2002} provide a detailed review of the Halo Model, so we will not go into any further details.

\subsection{Power Spectrum: Fitting Function}
The linear matter power spectrum can be calculated as:
\begin{eqnarray}
 \Delta_\textsuperscript{lin}^2(k_\textsuperscript{lin},z) = \frac{P_0  k_\textsuperscript{lin}^{3+n}}{2\pi^2} T^2(k_\textsuperscript{lin}) D_{+}^2(z).
\end{eqnarray}
where the transfer function, $T(k_\textsuperscript{lin})$, and the growing mode of linear perturbations, $D_{+}(z)$, can be approximated as described by \cite{Eisenstein1999}.

The non-linear power spectrum ($\Delta_\textsuperscript{nl}^2$) is known for extreme values of the wavenumber $k$ (i.e. at small $k$,
$\Delta_\textsuperscript{nl}^2 \simeq \Delta_\textsuperscript{lin}^2$, and at large $k$, $\Delta_\textsuperscript{nl}^2 \sim \Delta_\textsuperscript{lin}^3$), the latter derived using the so-called stable clustering \emph{ansatz}. In the light of these two
asymptotic regimes a fitting function was proposed to extrapolate $\Delta_\textsuperscript{nl}^2$ at all other $k$ directly from the
linear counterpart:
\begin{eqnarray}
 \Delta_\textsuperscript{nl}^2(k_\textsuperscript{nl},z) = \mathcal{F}[\Delta^2_{\textsuperscript{lin}}(k_\textsuperscript{lin},z)] ,
\label{eqnNlinpowspecdim}
\end{eqnarray}
where $k_\textsuperscript{nl} = k_\textsuperscript{lin} [1 + \Delta_\textsuperscript{nl}^2(k_\textsuperscript{nl},z) ]^{\frac{1}{3}}$. \cite{Peacock1996} found a fit for this for low density Universes with cold dark matter and a cosmological constant:
\begin{eqnarray}
  \mathcal{F}(x) = x \Big[\frac{1 + B\beta x +[Ax]^{\alpha\beta}}{1 + ([Ax]^\alpha g^3(z)/[Vx^{1/2}])^\beta }  \Big]^{1/\beta},
\end{eqnarray}
where,
\begin{eqnarray*}
    A      = 0.482(1 + n_\textsuperscript{eff}/3)^{-0.947},
\\  B      = 0.226(1 + n_\textsuperscript{eff}/3)^{-1.778},
\\  \alpha = 3.310(1 + n_\textsuperscript{eff}/3)^{-0.244},
\\  \beta  = 0.862(1 + n_\textsuperscript{eff}/3)^{-0.287},
\\  V      = 11.55(1 + n_\textsuperscript{eff}/3)^{-0.423},
\end{eqnarray*}
\begin{eqnarray*}
\textnormal{and } n_\textsuperscript{eff} = \frac{d \ln } { d \ln k_\textsuperscript{lin}}\Big( \frac{2\pi^2}{k_\textsuperscript{lin}^3}\Delta^2_\textsuperscript{lin}(k_\textsuperscript{lin})\Big) .
\end{eqnarray*}

This method can be used to approximate the non-linear dark matter power spectrum ($\Delta_{\rm DM}^2$). Under very general conditions, proved by \cite{Coles1993}, this can used to approximate the non-linear galaxy power spectrum, $\Delta_{\rm g}^2$, via a simple linear relationship of the form:
\begin{eqnarray}
 \Delta_{\rm g}^2(k,z) = b_\textsuperscript{eff}^2 \Delta_{\rm DM}^2(k,z).
\end{eqnarray}
We use an approximation for the bias proposed by \cite{Moscardini1998}, which is based on work by \cite{Mo1996} who suggested the following relation between the bias and mass density fluctuations:
\begin{eqnarray}
b(z,m) = 1 + \frac{1}{\delta_c} \Big[ \frac{ \delta_c^2}{\sigma_\textsuperscript{lin}^2(z,m)} -1 \Big ],
\end{eqnarray}
where $\delta_c$ is the critical linear density required for collapse and $\sigma_\textsuperscript{lin}^2$ is the linear variance in the smoothed density field. The effective bias can then be derived as,
\begin{eqnarray}
b_\textsuperscript{eff}(z) =\int\limits_{\ln(M_\textsuperscript{min})} ^\infty
b(z, m) n(z, m)d\ln m, \label{eq:MIH}
\end{eqnarray}
where $n$ is the normalized mass function, given by \cite{Sheth1999}. Note for simplicity we assume here that $M_\textsuperscript{min}$ is constant over redshift; given the larger redshift ranges a minimum mass varying with redshift may provide an improved fit but this approximation is sufficient for the comparison of models in this work.

Apart from the cosmological parameters, which we constrain to match other observations in this analysis, this fitting function method has just one free parameter (the minimum halo mass ${M_\textsuperscript{min}}$) whereas the Halo model has three: the minimum halo mass, the average mass of a halo with one satellite galaxy (M$_\textsuperscript{sat}$) and the slope of the first moment of the satellite galaxy HOD ($\alpha_s$).

\section{Fits to the Data}
\label{SECpoint2}
In this section, the models described above are compared with the observed angular correlation from recent \emph{Herschel} surveys and are used to put estimates on the minimum halo mass associated with these sources. The models we choose use current cosmological parameters from the Wilkinson Microwave Anisotropy Probe (WMAP) \citep[$\Omega_m$ = 0.274, $\Omega_\Lambda$ = 0.726, $\Omega_b$ = 0.046, h = 0.705, n = 0.96, $\sigma_8\textsuperscript{lin}$ = 0.812]{Komatsu2009}.
\begin{figure}
 \begin{centering}
  \includegraphics[width=80mm]{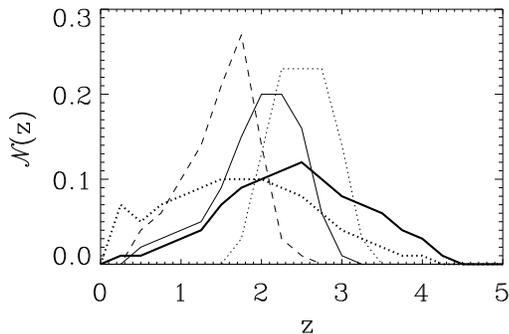}
  \caption{The normalized distribution of galaxies as a function of redshift for data sets from HerMES and H-ATLAS. HerMES estimates are from \citet{Cooray2010} and are shown in thin lines:  S$_{250}>$ 35mJy is a solid line,  S$_{350}$/S$_{250}>$ 0.85 is the dotted line and S$_{350}$/S$_{250}<$ 0.85 is the dashed line. H-ATLAS estimates are from \citet{Eales2010a} and are shown in thick lines: S$_{250} > 33$mJy is the dotted line, and S$_{350} > 35$mJy is the solid line. }
  \label{figNz}
 \end{centering}
\end{figure}

The redshift distribution of the sources plays a vital role in modeling the observed angular correlations. Figure \ref{figNz} shows the normalized redshift distributions, $\mathcal{N}(z)$, which were used in these calculations for both the HerMES and H-ATLAS data sets. The estimates for the redshift distribution for the HerMES data sets are consistant with those used by \cite{Cooray2010} in their analysis. Several estimates of the redshift distribution of objects in the H-ATLAS survey have been made. First, \cite{Amblard2010} used colour-colour diagrams to estimate the redshift distribution; however this method includes only a subset of the sources used in the angular correlation analysis and is known to be slightly biased towards higher redshift objects. Second, \cite{Dye2010} used optical counterparts of sub-mm sources to estimate the redshift distribution; but again this method only uses a subset of all the objects and is biased towards lower redshift objects. So far, the best estimates of the redshift distribution, in that they use all the available sources, are those by \cite{Eales2010a} who use Spectral Energy Distribution (SED) fitting to find a best fit redshift distribution. We use the estimates by \cite{Eales2010a} in this analysis, but the values for $\mathcal{N}(z)$ used here are \emph{for illustrative purposes only} and should not be regarded as definitive given the preliminary state of the data. We have tried a variety of alternative models and find that, for reasonable choices, the results for $\omega_\textsuperscript{obs}$ are not especially sensitive to a particular mean redshift in $\mathcal{N}(z)$; this is probably because of the relatively slow evolution of the power spectrum at low redshift in the concordance cosmology.  On the other hand the results are sensitive to the width of the distribution in $z$; the wider the distribution over $z$ the lower the amplitude of the angular correlation function. This is consistent with what you would expect if you consider that the clustering signal is more concentrated in narrower redshift bands.

\begin{figure}
 \begin{centering}
  \includegraphics[width=80mm]{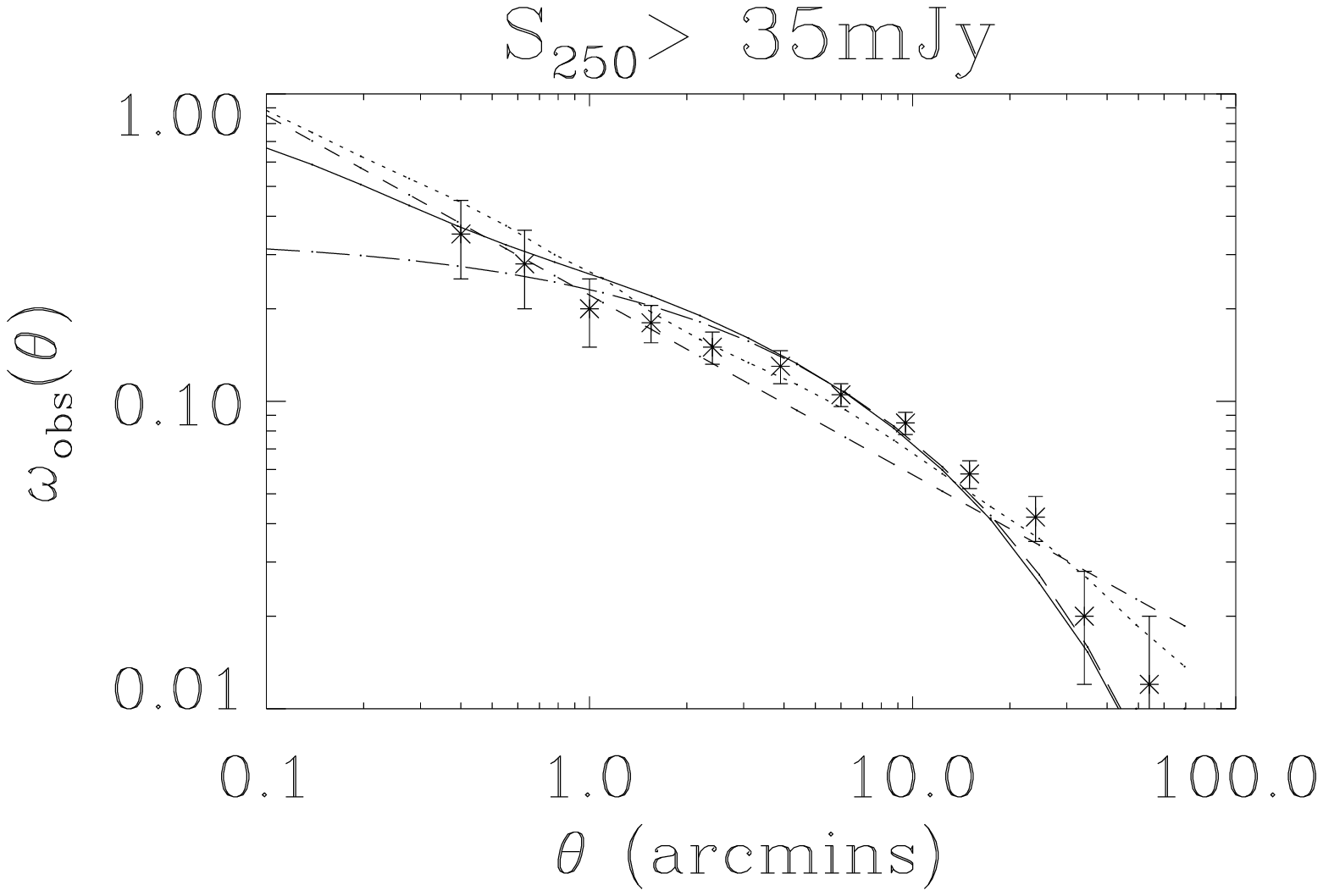}
  \includegraphics[width=80mm]{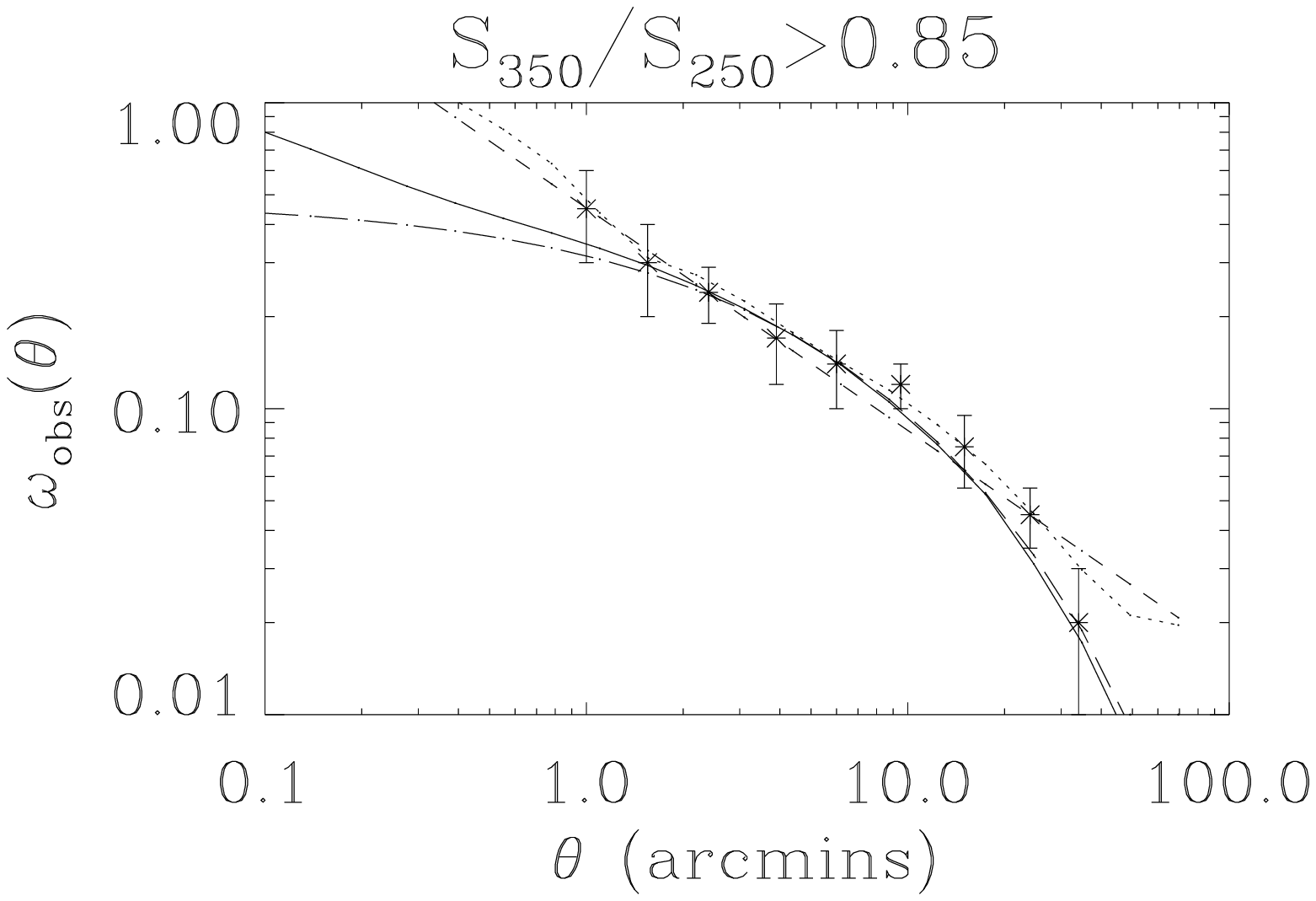}
  \includegraphics[width=80mm]{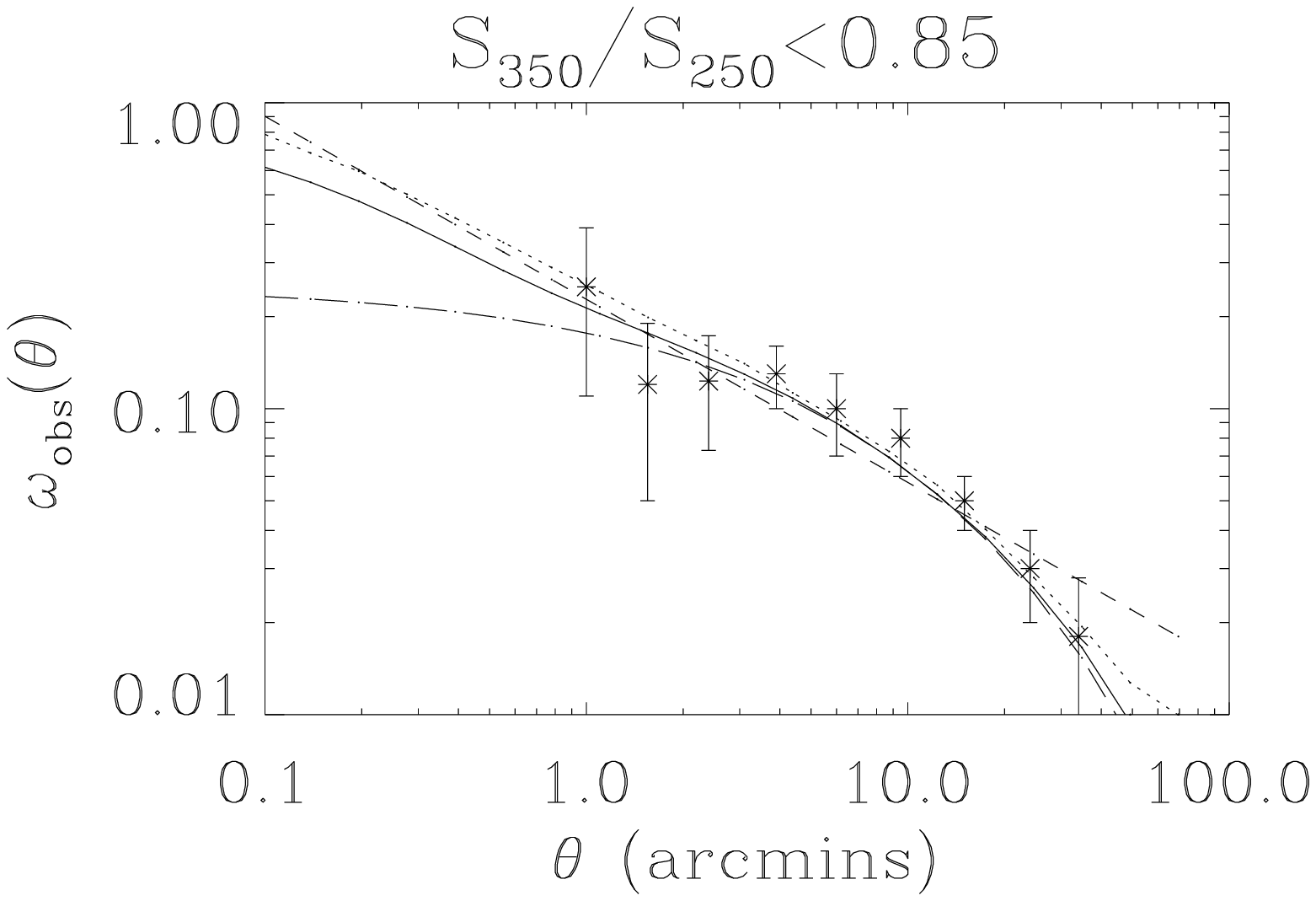}
  \caption{These plots show the angular correlation function $\omega_\textsuperscript{obs}(\theta)$ measured in the HerMES survey in the Lockman-SWIRE field. Sources were divided into three sets by flux density: S$_{250}>$ 35mJy (top),  S$_{350}$/S$_{250}>$ 0.85 (middle) and S$_{350}$/S$_{250}<$ 0.85 (bottom). Data is compared to four different theoretical models: the non-linear approximation using the Fitting Function (solid line), a power law (short dashed line), the Halo Model (dotted line), and the linear approximation (long dashed line). The Halo and Fitting Function methods fit the data well, as does the linear approximation.  The power law over estimates at large scales and underestimates at mid angular scales.}
 \label{figAng1}
 \end{centering}
\end{figure}

The \emph{Herschel} Multi-tiered Extragalactic Survey \citep[HerMES]{Oliver2010} will cover approximately 70 deg$^2$ of the sky in three different wavelength bands. The Lockman-SWIRE field, which is one of the shallower fields, was covered by the initial science run and results of the clustering have recently been published in \cite{Cooray2010} along with a Halo model analysis. Here we show just three of the available results, in Figure \ref{figAng1}, to compare with the theoretical models we have discussed above: S$_{250}>$ 35mJy, S$_{350}$/S$_{250}>$ 0.85 and S$_{350}$/S$_{250}<$ 0.85. From the observations we see that the angular correlation has similar amplitudes in each of the three data sets; the middle panel has the highest amplitude and this most likely because the redshift distribution of the component sources is the most narrow of the three data sets making the clustering signal more concentrated and appear stronger.

In Figure \ref{figAng1} four models are compared to the observations: the short dashed line shows a simple power law fit, the dotted line shows the fit from the Halo Model, the solid line shows the non-linear approximation using the Fitting Function approach, and the long dashed line shows just the linear spectrum. In the top two plots, the power-law fits the data well at low angular scales, but under estimates at mid scales and over estimates at higher scales. This is similarly the case for the last plot but it also overestimates clustering on small scales. The Halo model provides a better fit; the low and intermediate angular scales both fit well, although in the top two data sets the model is slightly overestimating at large angular scales. The fitting function approximation also fits the data well. The fit at intermediate scales is perhaps not quite as good as the Halo Model, but then this approach has fewer free parameters; the differences in the behavior at large angular scales is a consequence of slightly different best-fit values for the bias parameter. However, an important point to note here is that the \emph{linear} angular correlation function also provides a pretty good fit to each of the data sets. By comparing the linear and non-linear models we see that the transition between linear and non-linear regimes does not become
evident until quite small scales ($\sim$ 1 arcmin). The limited resolution of the \emph{Herschel} telescope makes it difficult to probe the clustering regime on scales much smaller than this.

In these models the amplitude of  large-scale clustering depends on the bias and that, in turns, boils down to the minimum halo mass through equation \ref{eq:MIH}. We can therefore use the measured clustering amplitude to get a rough estimate of the mass of halos hosting these galaxies. The values of the minimum halo mass corresponding to the theoretical models in Figure \ref{figAng1} are shown in Table \ref{TABmmin}. We see that for each of the data sets, the minimum halo masses found from both the Fitting Function and Halo models are in agreement (within the errors) and suggest a value of M$_\textsuperscript{min} \sim 10^{13.2 \pm 0.4} M_\odot h^{-1}$.
\begin{table}
  \begin{tabular}{|l||l|l||l|l|l|}
    \hline\hline
    Data Set & ${M^\textsuperscript{FF}_\textsuperscript{min}}$  & ${M^\textsuperscript{Halo}_\textsuperscript{min}}$  &  ${M_\textsuperscript{sat}}$  & $\alpha_s$  \\\hline
     S$_{250}>$ 35mJy          & 10$^{13.4}$ & 10$^{13.0}$  & 10$^{13.0}$ & $<$ 1.2\\
     S$_{350}$/S$_{250}>$ 0.85 & 10$^{13.2}$ & 10$^{13.2}$  & 10$^{13.4}$ & $<$ 1.2 \\
     S$_{350}$/S$_{250}<$ 0.85 & 10$^{13.4}$ & 10$^{13.0}$  & 10$^{13.2}$ & $>$ 1.4 \\
    \hline
  \end{tabular}
    \caption{Best fit parameters for the Halo and Fitting Function models for the HerMES results. M$_\textsuperscript{min}$ is the minimum halo mass for which galaxies can form, M$_\textsuperscript{sat}$ is the average mass of a halo with one satellite galaxy and $\alpha_s$ is the slope of the first moment of the satellite galaxy HOD.  All masses are in units $M_\odot h^{-1}$. We see that the M$_\textsuperscript{min}$ found for each data set are consistent between models, given an accuracy of 10$^{\pm 0.2}$ $M_\odot h^{-1}$ on the masses.}
  \label{TABmmin}
\end{table}
The results using our Halo model are consistent with those found in the Halo Model analysis by \cite{Cooray2010}.
\begin{figure}
 \begin{centering}
  \includegraphics[width=80mm]{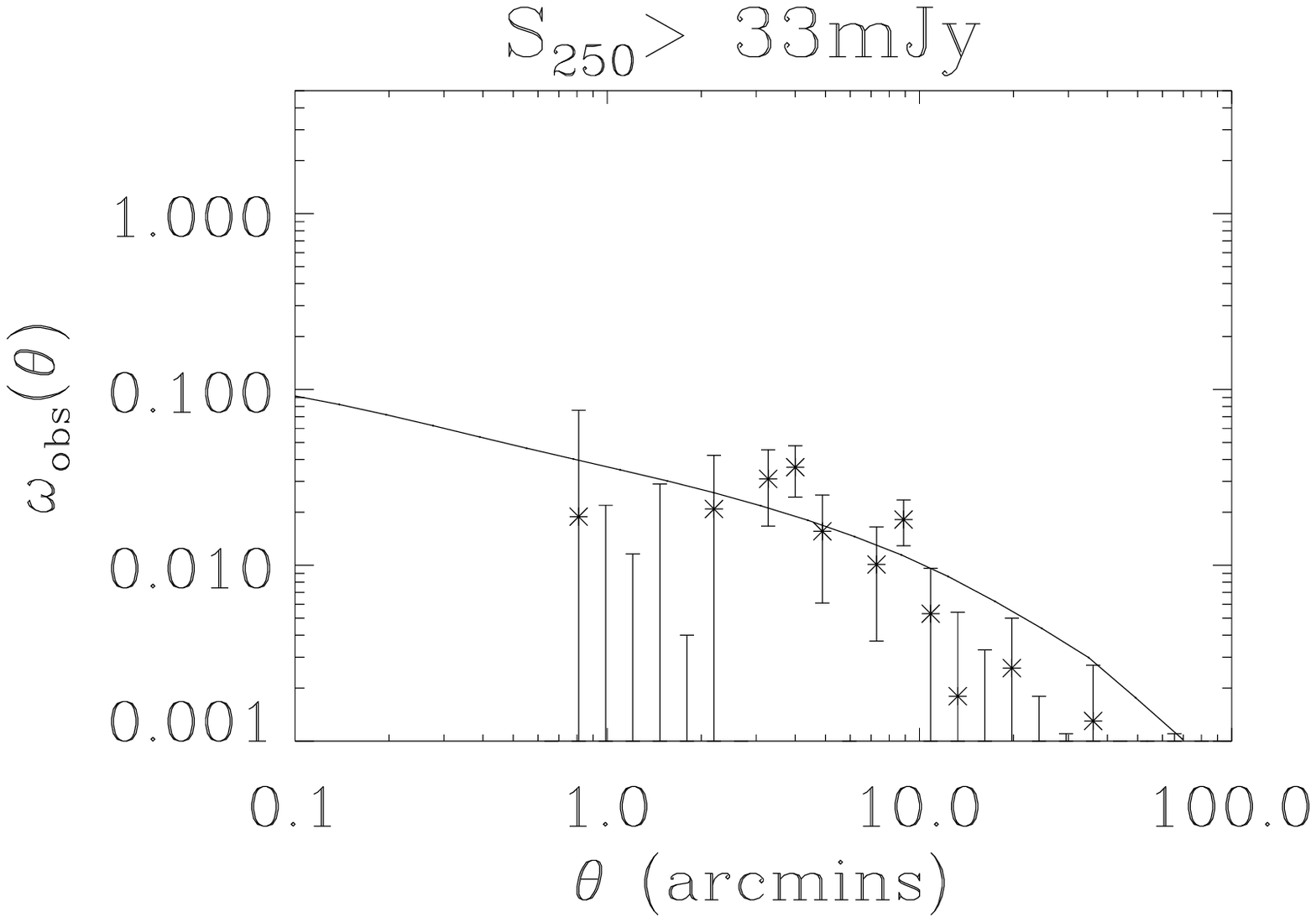}
  \includegraphics[width=80mm]{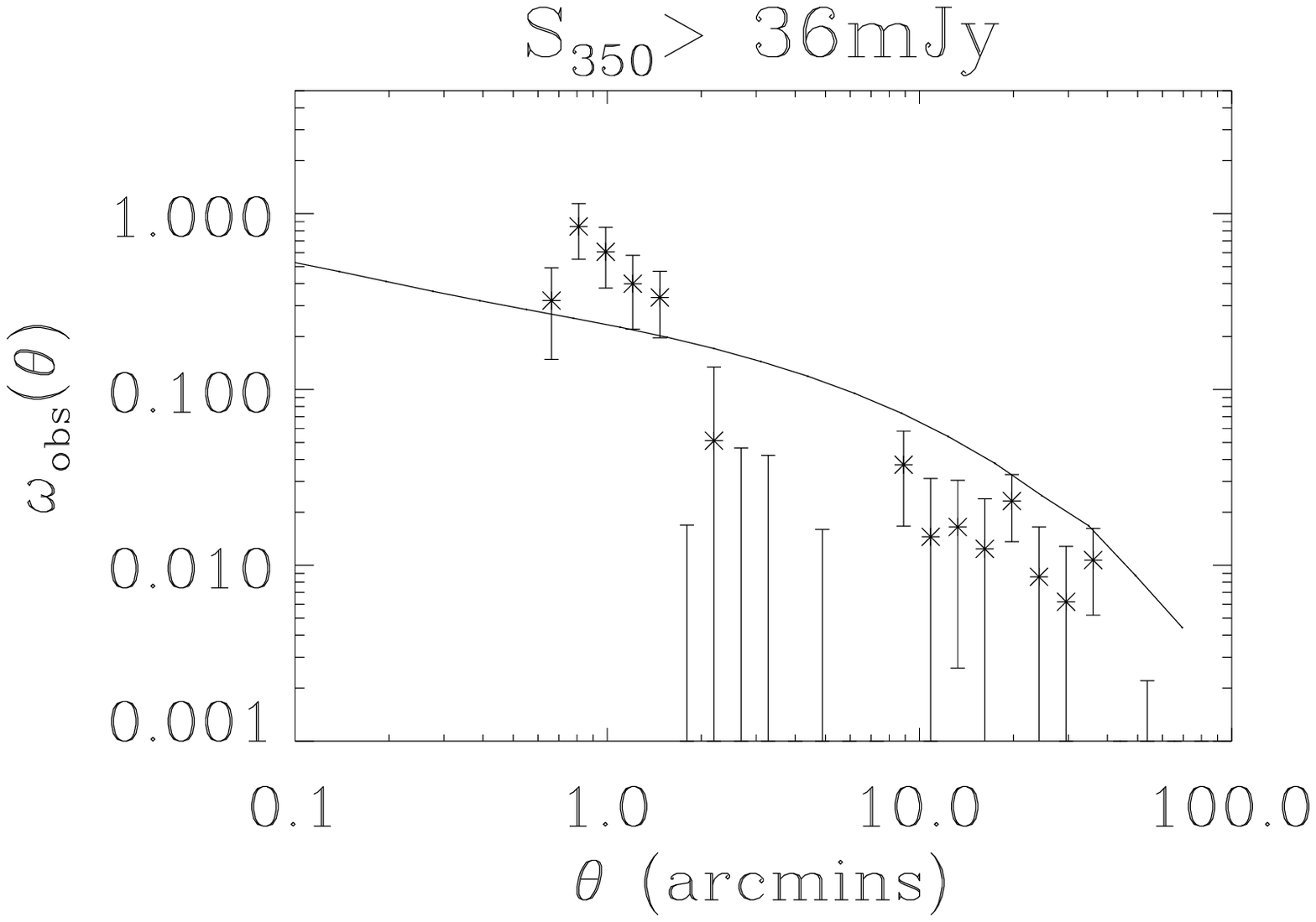}
  \caption{These plots show the angular correlation function $\omega_\textsuperscript{obs}(\theta)$ measured in the science demonstration phase of the H-ATLAS survey. Sources were divided into sets by flux density: S$_{250} > 33$mJy (top) and S$_{350} > 36$mJy (bottom). The data are compared to the Fitting Function model (solid line). }
  \label{figAng2}
 \end{centering}
\end{figure}

The other results we have analyzed are from the \emph{Herschel} ATLAS survey \citep[H-ATLAS]{Eales2010} which will cover 550 deg$^2$ of the sky, in 5 different wavebands covering the far infrared to sub-millimetre. The science demonstration phase of the survey has just been completed (which covers roughly 1/30 of the final H-ATLAS data-set) and \cite{Maddox2010} have released measurements of the angular correlation function of the galaxies observed. Figure \ref{figAng2} shows different flux cuts for S$_{250} > 33$mJy (top) and S$_{350} > 36$mJy (bottom). The results are quite noisy so none of the models fit particularly well - a detailed analysis will have to wait until the completion of the survey. This noise is thought to be due to galactic cirrus in the H-ATLAS data as the number of detections are very similar to the HerMES data sets and both surveys are similarly confusion limited. In this case, therefore we have just plotted the Fitting Function results against the data. It was used to find approximate values of the corresponding minimum halo masses, which are 10$^{12.5}$ and 10$^{13.2}$ $M_\odot h^{-1}$ for the examples plotted in Figure \ref{figAng2} top to bottom respectively.  The scale of the $\omega_\textsuperscript{obs}$ results for the S$_{250} > 33$mJy flux cut is significantly smaller that those in the other examples. This is predominately due to the wide bimodal redshift distribution (see Figure \ref{figNz}) although it does still suggest a slightly lower minimum mass $\sim$ 10$^{12.5}$ $M_\odot h^{-1}$. This minimum mass is consistent with that found in the cross correlation analysis by \cite{Guo2010}.

\section{Discussion and Conclusions}
\label{SECcon}
In this paper, we have compared models of the angular correlation function to data from the science demonstration phase of \emph{Herschel}. We highlight a Fitting Function method which provides an improved fit to the data than a power law, and similar to that of the Halo model. It has just one free parameter, the minimum halo mass, compared to the two and three for a simple power-law and the halo model respectively. The halo mass is more meaningful physically as a parameter than those involved in the power law fit, so the Fitting Function is a much better method than the power-law for a quick-and-simple analysis. Although neither as sophisticated nor as flexible as the Halo model, it remains a useful tool that is perfectly adequate for modelling currently available data. For example, the minimum halo mass found using our Fitting Function model is consistent with that found using the Halo Model.

Another point of interest is that, in fact, the linear angular correlation function also provides a reasonable fit to the data. Owing to the limited resolution of the \emph{Herschel} telescope it is difficult to identify pairs of galaxies sufficiently close together on the sky to probe anything but the mildly non-linear regime. The currently available data provide some evidence of a transition between the linear and non-linear regimes, but they provide no unambiguous detection of the change of shape in $\omega_\textsuperscript{obs}$ that the Halo Model predicts. This does not mean the Halo Model is incorrect, of course, but what it does mean is that, at least for the time being, the paraphernalia involved in modelling intra-halo correlations is rather superfluous for  these objects; simpler models can yield perfectly adequate results.

Finally, we stress that the data sets to which we have applied these models are preliminary. The biggest stumbling-block to a more complete analysis relates to the considerable uncertainties in the redshift distribution $N(z)$ of the sources involved. The choices we adopted for this analysis are for illustration only so the results should not be regarded as definitive. Further data, especially ancillary data providing spectroscopic redshifts, will be needed before the precise nature of these galaxies can be determined.

\section*{Acknowledgments}
Jo Short receives funding from an STFC studentship. For the purposes of this work Peter Coles is supported by STFC Rolling Grant ST/H001530/1. We thank Steve Eales, Loretta Dunne and Steve Maddox for helpful discussions, and are grateful to them for allowing us to
use data from Eales et al. (2010a) in advance of publication.

\end{document}